\documentclass[journal,10pt]{IEEEtran}

\usepackage[utf8]{inputenc}
\usepackage{amsfonts}
\usepackage{amssymb}
\usepackage{amsmath}
\usepackage{microtype}
\usepackage{graphicx}
\usepackage{booktabs}
\usepackage{hyperref}
\usepackage{multirow}
\usepackage{footnote}
\usepackage{tablefootnote}
\usepackage{siunitx} 
\usepackage{xcolor} 
\usepackage{gensymb} 
\usepackage{commath} 
\usepackage{bm} 
\usepackage[noadjust]{cite} 
\usepackage{environ} 
\usepackage{breqn} 

\makesavenoteenv{table}
\makesavenoteenv{table*}
\makesavenoteenv{tabular}

\newcommand{\R}[1] {\textnormal{#1}}

\NewEnviron{NORMAL}{%
    \scalebox{1}{$\BODY$} 
} 

\DeclareMathOperator*{\E}{\mathbb{E}}

\begin{document}
 
\title{MIMO-SAR: A Hierarchical High-resolution Imaging Algorithm for mmWave FMCW Radar in Autonomous Driving}

\author{Xiangyu Gao, Sumit Roy, \IEEEmembership{Fellow,~IEEE}, Guanbin Xing, \IEEEmembership{Member,~IEEE}

\thanks{X. Gao, S. Roy are with the Dept. of Electrical and Computer Engineering, University of Washington, Seattle, WA, 98195, USA. E-mail: xygao@uw.edu, sroy@uw.edu. 
G. Xing was with the Dept. of Electrical and Computer Engineering, University of Washington, Seattle, WA, 98195, USA. E-mail: gxing@uw.edu.}}



\maketitle

\begin{abstract}
Millimeter-wave radars are being increasingly integrated into commercial vehicles to support advanced driver-assistance system features. A key shortcoming for present-day vehicular radar imaging is poor azimuth resolution (for side-looking operation) due to the form factor limits on antenna size and placement. 
In this paper, we propose a solution via a new multiple-input and multiple-output synthetic aperture radar (MIMO-SAR) imaging technique, that applies coherent SAR principles to vehicular MIMO radar to improve the side-view (angular) resolution. The proposed 2-stage hierarchical MIMO-SAR processing workflow drastically reduces the computation load while preserving image resolution. To enable coherent processing over the synthetic aperture, we integrate a radar odometry algorithm that estimates the trajectory of ego-radar. The MIMO-SAR algorithm is validated by both simulations and real experiment data collected by a vehicle-mounted radar platform.
\end{abstract}


\section{Introduction}

\IEEEPARstart millimeter-wave (mmWave) radars potentially provide highly accurate object detection and localization (range, velocity and angle) independent of environmental conditions \cite{8828025}. Thus, they are being increasingly integrated into commercial vehicles as sensors for environmental perception in future (semi) autonomous driver assistance operational modes. In particular, side-looking radars are widely used to support the lane change or keeping assist, blind spot detection, and rear cross-traffic alert. For these applications, a high-resolution radar image is key to effective separation of close objects, detection of the spatial extension of traffic participants, and enhanced object recognition \cite{ramp}. 

\SI{77}{GHz} frequency modulated continuous wave (FMCW) radars can already achieve centimeter-level range resolution ($\sim$ \SI{3.75}{cm} with \SI{4}{GHz} sweep bandwidth). The remaining challenge is to improve the azimuth angle (cross-range) resolution. There exist two main classes of approaches for side-view automotive radars to improve azimuth resolution: multiple-input and multiple-out (MIMO) processing \cite{ti_mimo, gao2019experiments, 8546566, gao2021perception}, and synthetic aperture radar (SAR) \cite{9104293, glob, 8443554}. 

\begin{figure*}
    \centering
    \includegraphics[width=7.0in, trim=1 3 1 1,clip]{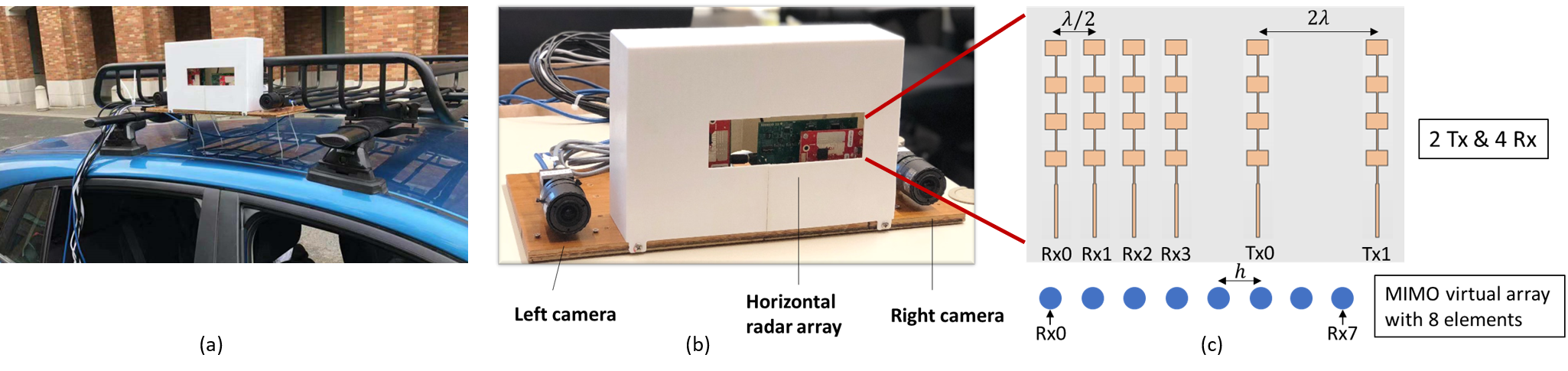}
    \caption{(a) Integrated radar-camera data capture platform mounted on one vehicle; (b) FLIR cameras and TI radar board with 1-D horizontal array; (c) Horizontal antenna array includes 2 Tx with distance $2 \lambda$ and 4 Rx with distance $\lambda / 2$, yielding virtual MIMO array with 8 Rx elements.}
  \label{platform}
\end{figure*}

MIMO radar systems transmit mutually orthogonal signals from multiple transmit antennas, followed by {\em joint} processing of the signals at all receive antennas \cite{ti_mimo} for extraction of target information. For example, if a MIMO radar has $N_{\R{Tx}}$ transmit antennas (Tx) and $N_{\R{Rx}}$ receive antennas (Rx) with appropriate arrangement, a $N_{\R{Tx}} N_{\R{Rx}}$-element virtual antenna array \cite{ti_mimo} can be created by conducting beamforming on the received signals, thereby obtaining a finer azimuth resolution compared with its phased array counterpart. As shown in Fig.~\ref{platform}(c), the state-of-art Texas Instrument (TI) 77 GHz FMCW radar chip AWR1843 is capable of synthesizing 8 virtual receivers with 2 Tx and 4 Rx \cite{ti1843evm} using time division multiplexing (TDM) \cite{ti_mimo, 8645667}, resulting in \ang{15} azimuth resolution \cite{gao2019experiments} which is insufficient for high-resolution target discrimination.

SAR is a well-known imaging technique that overcomes the limits of small physical aperture on angular resolution by coherently processing the returns from a series of transmitted pulses to create a large {\em synthetic} aperture\cite{piofmd}. SAR is applicable for imaging stationary targets and background with active source mounted on a moving platform (e.g. vehicle mounted radar), while the moving targets can be observed via inverse SAR (ISAR) concepts with a stationary radar setup \cite{9048497, Kulpa2013ExperimentalRO}. Besides, automotive (vehicle-based) SAR algorithms need to cope with specific difficulties - notably estimation of vehicle motion and (importantly) achieving real-time operation. Accordingly, prior research on automotive SAR algorithms has focused on speeding up post-processing via techniques to reduce the algorithmic run-time complexity \cite{9104293, 8546620, 6982206, 6960071}.

In this paper, we propose a new MIMO-SAR algorithm that exploits key features of FMCW MIMO radar to achieve computationally efficient SAR imaging. Specifically, MIMO processing is used for initial low-cost target detection and localization to narrow down the {\em region of interest} (ROI) for subsequent finer-resolution SAR processing. We adopt the time-domain backprojection SAR algorithm \cite{glob} - that lends itself naturally to parallel processing with graphics processing units (GPU) \cite{sar_tbx, 6510489} - to progressively operate on ROI as new snapshots are coherently added and processed. To reduce SAR processing frequency (i.e. PRF) and consequent complexity, the returns from a train of FMCW chirps are stored in a range-velocity-angle (RVA) data cube that is processed via the 3D-FFT algorithm (Section \ref{sec_3dfft}); thereafter we select the max-intensity velocity component for subsequent SAR processing. Above 2-stage hierarchical workflow drastically reduces the computation load while preserving high-resolution imaging. 

Coherent SAR processing requires phase compensation of source motion trajectory to achieve coherent in-phase processing. In turn, this requires accurate vehicle ego-motion estimation. The traditional solution is to use on-board inertial measurement units (IMU) \cite{kong2000inertial} that combines measurements from the wheel speed sensor, gyroscopes, and accelerometers. However, high-precision IMUs are cost-prohibitive for automotive applications, inspiring the need for self-contained alternatives such as radar odometry, to determine the velocity and direction of motion of the vehicular radar \cite{6728341, 9096265, 8995552}. Our approach is based on analyzing the distribution of the radial velocities of the received reflections (targets) over their azimuth angles, which can be provided by the MIMO processing stage described above.

In summary, the main novel contributions of this paper are three-fold:
\begin{itemize}
\item A new hierarchical MIMO-SAR algorithm that reduces computational complexity while preserving high-resolution imaging;
\item A radar odometry algorithm to estimate the trajectory of ego-radar and enable MIMO-SAR coherent processing.
\item Validation of the proposed MIMO-SAR algorithm by both simulations (Section \ref{simu res}) and real data (see Fig.~\ref{platform} and Section \ref{exper res}).
\end{itemize}

The rest of the paper is organized as follows. We summarize the model for FMCW MIMO radar processing in Section \ref{FMCW model}, and present the new MIMO-SAR imaging algorithm in Section \ref{MIMO SAR algo}. The requirements for enhanced MIMO-SAR system design is summarized in Section \ref{sys para requr}. The simulation and experiment results are shown in Section \ref{simu res} and Section \ref{exper res} respectively. 

\section{FMCW Signal Model for MIMO Radar \label{FMCW model}}

\subsection{Signal Model}

\begin{figure}
    \centering
    \includegraphics[width=3.2in]{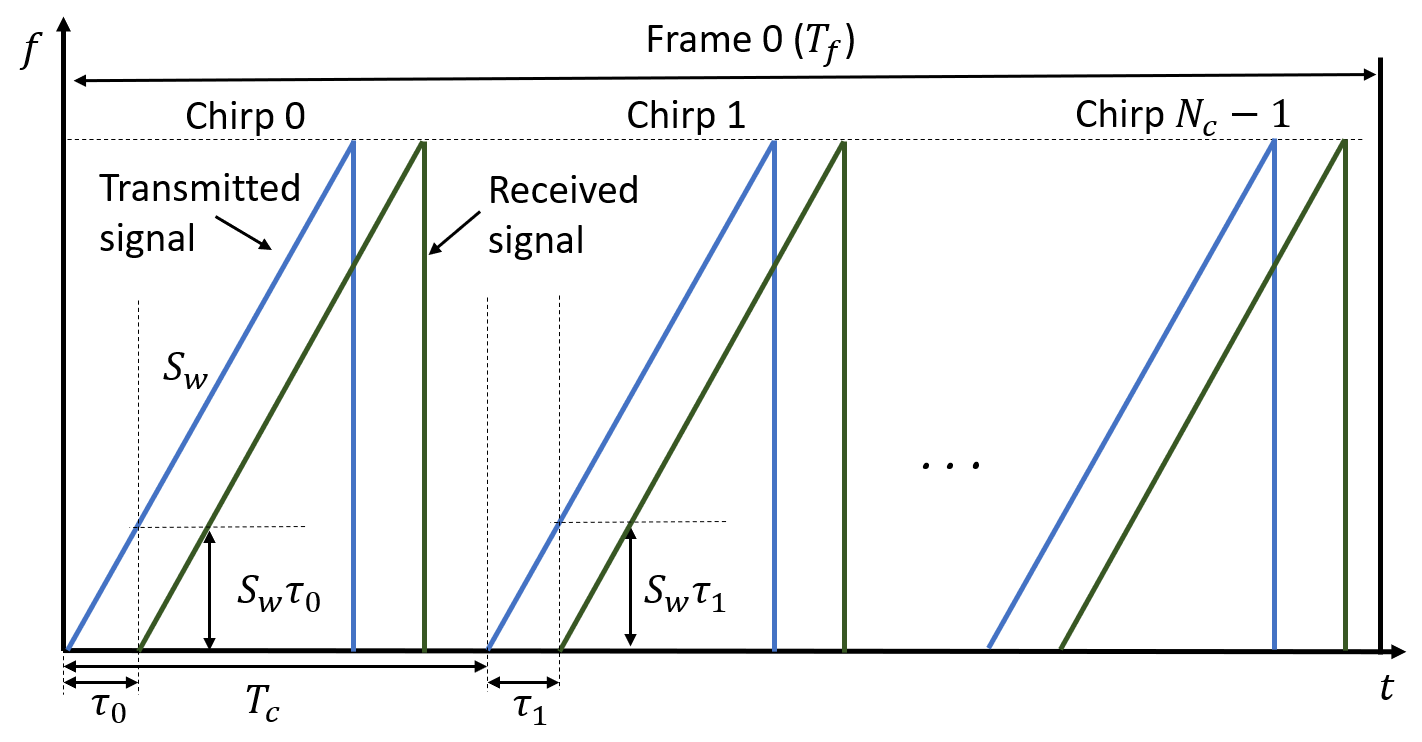}
    \caption{The transmitted FMCW chirps and corresponding return signal.}
    \label{chirp}
\end{figure}

The FMCW source transmits a sequence of $N_{\R{c}}$ chirps (with chirp duration $T_{\R{c}}$) that constitute a frame of duration $T_{\R{f}}$, as shown in Fig.~\ref{chirp}. With carrier frequency $f_{\R{c}}$, sweep slope $S_{\R{w}}$, amplitude $A_{\R{T}}$, and initial phase $\phi_0$, the FMCW transmit signal for chirp $0$ is given by \cite{4418542}:
\begin{equation}
     s_{\R{T}} (t)=A_{\R{T}}  \cos \left(2\pi \left(f_{\R{c}} t+\frac{1}{2}S_{\R{w}} t^2 \right) + \phi_0 \right) 
\label{tx}
\end{equation}

\begin{figure}
\centering
\includegraphics[width=3.3in, trim=2 3 2 2,clip]{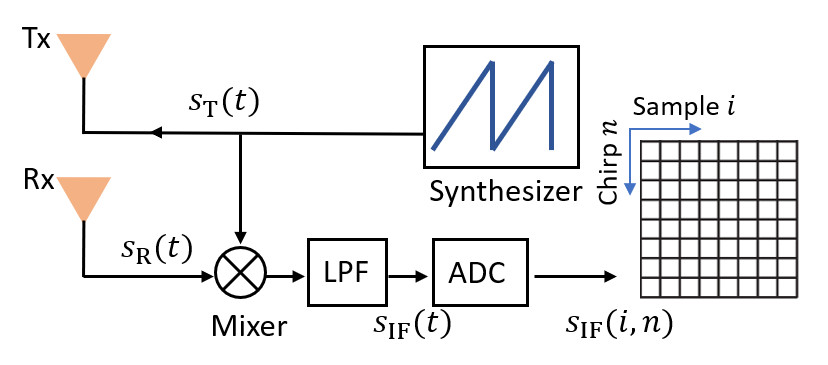}
\caption{The receiver chain for FMCW radar.}
  \label{receiver}
\end{figure}

For a target at distance $r$ from radar, the delayed return of the transmitted signal corresponding to the round-trip delay $\tau=\frac{2r}{c_0}$ is given by $s_{\R{R}}(t) = s_{\R{T}}(t-\tau)$, where $c_0$ is the speed of light. In the receiver chain shown in Fig.~\ref{receiver}, the received signal $s_{\R{R}}(t)$ is mixed with the transmitted signal $s_{\R{T}}(t)$ to obtain the de-chirped signal which is applied to a low-pass filter to filter out the high-frequency component and retain the difference signal \cite{udmmw}. The resulting intermediate frequency (IF) signal \cite{essay70986} corresponding to chirp $0$ is given by:
\begin{dmath}
    s_{\R{IF}}(t) = \frac{A_{\R{T}} A_{\R{R}}}{2} \Bigg\{ \cos \left[2 \pi\left(S_{\R{w}} \tau t+f_{\R{c}} \tau-\frac{1}{2} S_{\R{w}} \tau^{2}\right)\right] \Bigg\}
 \label{sif}
\end{dmath}

\begin{figure}
\centering
\includegraphics[width=3in, trim=2 3 2 2,clip]{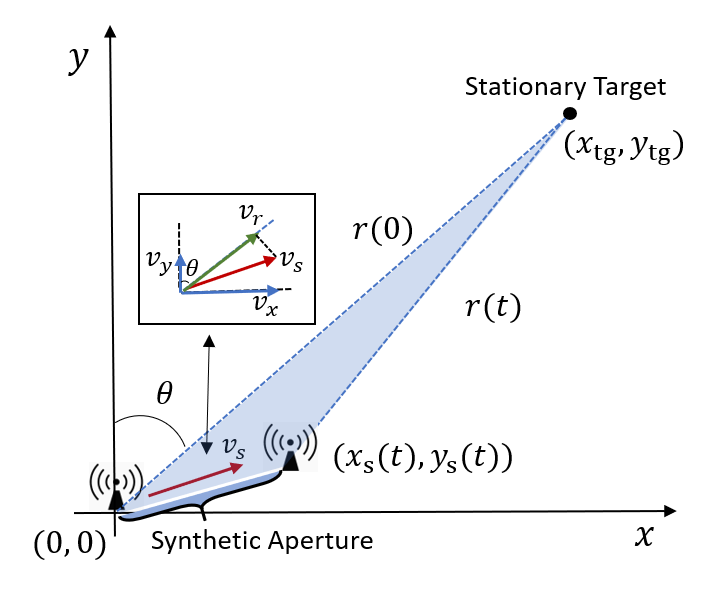}
\caption{The illustration of the scenario for a  moving radar within one frame.}
  \label{sentar}
\end{figure}

Consider the scenario shown in Fig.~\ref{sentar}, where the vehicle-mounted radar is assumed to have a constant velocity $\bm{v_{\R{s}}}=(v_{\R{x}},v_{\R{y}})$ relative to the stationary target within a frame \footnote{We assume the vehicle-mounted radar motion is linear over the (short) frame $T_{\R{f}}$ \cite{8995552}. For different frames, the velocity can be different, as when the trajectory is non-linear as shown in Fig.~\ref{def_reg}.}, and $v_{\R{x}}, \, v_{\R{y}}$ are the velocity components along x- and y-direction. The radar is assumed located at the origin at $t=0$ without loss of generality; at time $t$, the distance between radar location $(x_{\R{s}}(t), y_{\R{s}}(t))$ and that of stationary target ($x_{\R{tg}}, y_{\R{tg}}$) is given by $r(t) = \sqrt{(v_{\R{x}} t - x_{\R{tg}})^2 + (v_{\R{y}} t - y_{\R{tg}})^2}$.

For very short chirp duration $T_{\R{c}}$ (a few microseconds), the round-trip delay is assumed fixed for each chirp and therefore can be represented as $\tau_n = \frac{2r(nT_c)}{c_0} $ for chirp $n$. Across chirps within a frame, $\tau_n$ can be expressed as a linear approximation with the relative radial velocity $v_{\R{r}}$ (see Fig.~\ref{sentar}, where $v_{\R{r}}$ is velocity component projected onto the radial direction $\theta$) \cite{aut_rad_review}, i.e.,
\begin{equation}
    \tau_n = \tau_{0} - \frac{2 v_{\R{r}} n T_{\R{c}}}{c_0}
\label{tau}
\end{equation}

\noindent where $\tau_0 = \frac{2\sqrt{x_{\R{tg}}^2 + y_{\R{tg}}^2}}{c_0}$ is the round-trip delay for chirp $0$.

For above scenario, the quadrature analog-to-digital conversion (ADC) samples of the IF signal of chirp $n$ is given by $s_{\R{IF}}(i,n)$, where $i$ is the index of ADC samples within a chirp (see Fig.~\ref{receiver}). The samples $s_{\R{IF}}(i,n)$ are obtained by replacing $t$ and $\tau$ in \eqref{sif} with $i / f_{\R{s}}$ and $\tau_n$, where $f_{\R{s}}$ is the ADC sampling frequency \cite{aut_rad_review}.
\begin{dmath}
    s_{\R{IF}}(i,n)=\frac{A_{\R{T}} A_{\R{R}}}{2} \exp \left(j 2 \pi\left(S_{\R{w}} \tau_n \frac{i}{f_{\R{s}}} + f_{\R{c}} \tau_n - \frac{1}{2} S_{\R{w}} \tau_n^{2}\right)\right)
    \label{ifs}
\end{dmath}


As mentioned earlier, for the MIMO radar with $N_{\R{Tx}}$ transmitters and $N_{\R{Rx}}$ receivers, we can create a $N_{\R{Tx}} N_{\R{Rx}}$-element virtual array via transmitting orthogonal waveforms (see Fig.~\ref{platform}(c)). Therefore, the ADC-sampled IF signal \eqref{ifs} is extended for virtual MIMO operation as follows: for Rx $q$ of virtual array, we add a phase term 
$2 \pi q h\sin{\theta} / \lambda$ corresponding to the far-field target's angle of arrival $\theta$ \cite{piofmd, 526899} into \eqref{ifs}.
\begin{dmath}
    s_{\R{IF}}(i, n, q)=\frac{A_{\R{T}} A_{\R{R}}}{2} \exp \left(j 2 \pi\left(S_{\R{w}} \tau_n \frac{i}{f_{\R{s}}} + f_{\R{c}} \tau_n - \frac{1}{2} S_{\R{w}} \tau_n^{2} + \frac{q h\sin{\theta}}{\lambda} \right)\right)
 \label{sifmimo}
\end{dmath}

\noindent where $h$ is inter-Rx distance of virtual array ($h=\lambda/2$ in Fig.~\ref{platform}(c)).

\subsection{Range, Velocity and Angle Estimation \label{sec_3dfft}}
We operate the 3-dimensional fast Fourier transform (\textbf{3D-FFT}) - Range FFT, Velocity FFT, and Angle FFT - on the digitized IF signal \eqref{sifmimo} to estimate the range, Doppler velocity, and azimuth angle spectrums. The obtained 3D spectrum is named Range-Velocity-Angle \textit{RVA} cube.

\subsubsection{\textbf{Range Estimation}}
The IF signal has a beat frequency $f_{\R{b}}=S_{\R{w}} \tau_n$, where $\tau_n$ is related to the distance to target. To estimate the beat frequency, a fast Fourier transform ({\em Range FFT}) is used to convert the time domain IF signal into the frequency domain \cite{gao2019experiments}; the peaks in resulting spectrum (or range profile) can be transformed to the distance of target.
The Range FFT implemented on chirp $n$ and receiver $q$ is given by $S_{\R{R}}(m_{\R{r}}, n, q) = \mathcal{F}\{ s_{\R{IF}}(i, n, q) \}$, where $m_{\R{r}}$ is the range bin. The range resolution is determined by the swept RF bandwidth $B$ with the well-known equation $R_{\R{res}}=\frac{c_0}{2B}$.
\vspace{0.4em}


\subsubsection{\textbf{Velocity Estimation}}

According to \eqref{tau} and \eqref{ifs}, the relative radial velocity $v_{\R{r}}$ will cause a Doppler phase shift $\Delta \phi_{\R{v}} = \frac{4\pi v_{\R{r}} T_{\R{c}}}{\lambda}$ in IF signal between consecutive chirps. Hence, a fast Fourier transform ({\em Velocity FFT}) is executed across chirps to estimate phase shift and then transform it to velocity \cite{gao2019experiments}.
The Velocity FFT performed on the range profile is expressed as $S_{\R{RV}}(m_{\R{r}}, m_{\R{v}}, q) = \mathcal{F}\{ S_{\R{R}}(m_{\R{r}}, n, q) \}$, where $m_{\R{v}}$ is velocity bin. The velocity resolution of this method is given by $V_{\R{res}}=\frac{\lambda}{2N_{\R{c}} T_{\R{c}}}$ \cite{iovescu2017fundamentals}, where $N_{\R{c}}$ is number of chirps in a frame.
\vspace{0.4em}


\subsubsection{\textbf{Angle Estimation} \label{angle_estimat}}
 
The return from a target located at angle $\theta$ (far field) results in a steering vector with fixed phase shift $\Delta \phi_{\theta} = \frac{2 \pi h \sin{\theta}}{\lambda}$ for a uniform linear array \cite{526899}. Then the angle estimation can be conducted by a fast Fourier transform ({\em Angle FFT}) across the signal over the Rx elements \cite{ti_mimo, gao2019experiments}.
The Angle FFT is represented as $S_{\R{RVA}}(m_{\R{r}}, m_{\R{v}}, m_\theta) = \mathcal{F}\{ S_{\R{RV}}(m_{\R{r}}, m_{\R{v}}, q) \}$, where $m_\theta$ is azimuth angle bin. The angle resolution for MIMO radar is $\theta_{\R{res}}=\frac{\lambda}{N_{\R{Tx}}N_{\R{Rx}}h\cos{\theta}}$ \cite{iovescu2017fundamentals}.

For non-stationary targets, the motion-induced phase errors should be compensated before Angle FFT on the virtual Rx elements corresponding to the second Tx in case of TDM-MIMO \cite{ti_mimo}. According to \cite{8052088}, these are corrected via phase compensation of $\frac{\Delta \phi_{\R{v}}}{2}$ (half the estimated Doppler phase shift), which can be obtained from the Velocity FFT results.



\begin{figure*}
  \centering
  \includegraphics[width=7.0in, trim=2 3 2 2,clip]{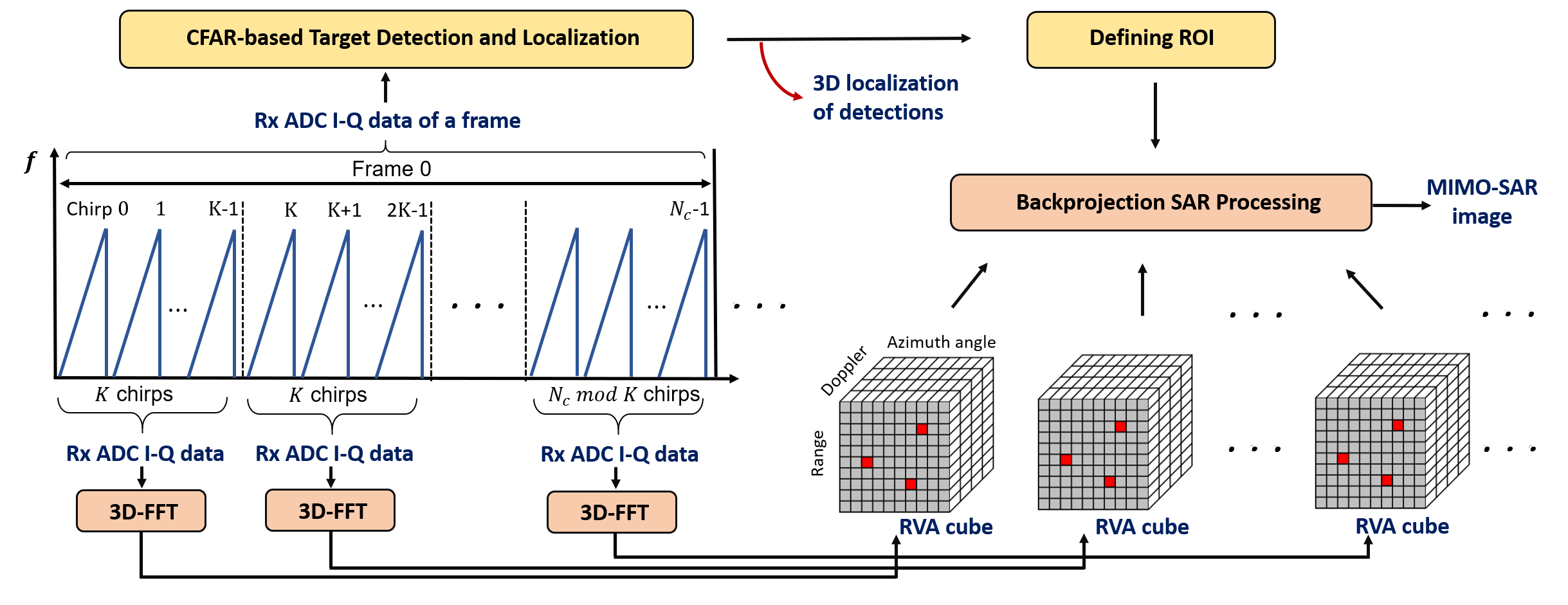}
  \caption{MIMO-SAR algorithm illustration: target detection and localization (MIMO processing) are operated for every frame; backprojection SAR algorithm is performed on RVA cubes generated from every $K$ chirps by 3D-FFT.}
  \label{system}
\end{figure*}

\begin{figure*}
  \centering
  \includegraphics[width=7.1in, trim=1 3 1 1,clip]{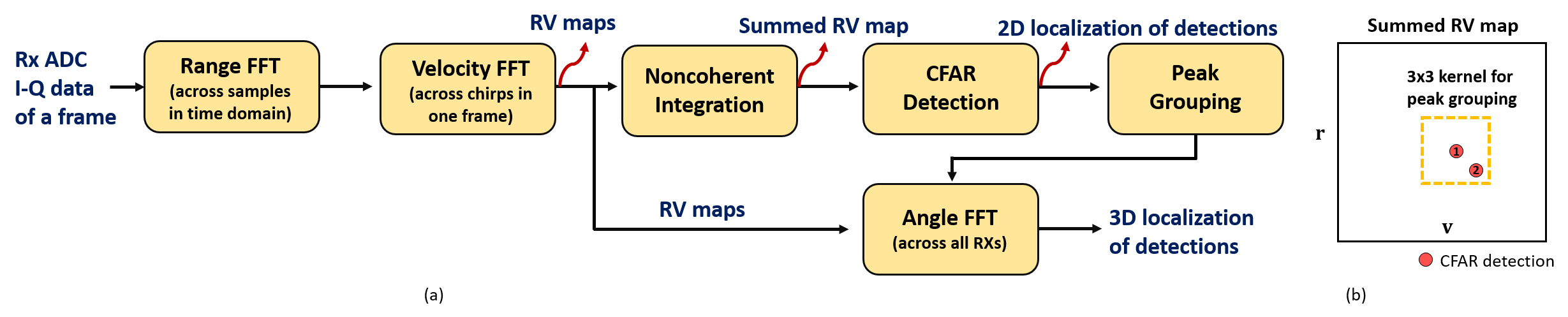}
  \caption{(a) The block diagram for CFAR-based target detection and localization algorithm; (b) Illustration for peak grouping.}
  \label{cfar_target_det}
\end{figure*}

\section{MIMO-SAR Algorithm \label{MIMO SAR algo}}

The hierarchical MIMO-SAR algorithm workflow is divided into 2 stages - MIMO processing, and SAR processing - to reduce the computation load, as shown in Fig.~\ref{system}. The load decrease is achieved by both spatially focusing on the Regions-of-Interest (ROI) provided by MIMO processing and temporally reducing SAR sampling rate (or PRF) using RVA cubes. Specifically, MIMO processing is performed for every frame to localize the detected targets in the field of view such that we can narrow down the ROI for subsequent SAR imaging. For SAR processing, we implement the backprojection algorithm \cite{glob, sar_tbx} on the RVA data cubes (provided by 3D-FFT on every $K$ chirps \footnote{The MIMO localization and radar odometry are operated every frame because the accuracy and resolution of velocity estimation is greater with longer observation \cite{iovescu2017fundamentals}. However, SAR image is updated once for every K chirps (a subset of frame) to avoid the azimuth aliasing.}) to obtain high-resolution imaging. 

\subsection{MIMO Processing \label{mimo_region}}

 MIMO radar can provide estimation of range, Doppler (radial) velocity, and azimuth angle for the detected targets in the field of view, which confers two advantages to our MIMO-SAR algorithm compared to traditional SAR processing: 1) Initial target localization enables a hierarchical approach whereby selecting ROI within imaging plane for subsequent SAR processing effectively reduces computation complexity; 2) Analyzing estimated Doppler velocities and azimuth angles for detected stationary targets enables radar ego-motion estimation, which is necessary for SAR phase compensation along trajectory (Section \ref{rad odo}). 

The computation load of SAR imaging increases in direct proportion to the potential target area of the SAR image plane. Classic SAR algorithms coherently process each cell in the SAR image plane. This is redundant since in practice, most of the energy (or useful information) is concentrated in a subset. Therefore, to reduce the computational complexity, we propose to select the ROIs in SAR image based on the localization of detections provided by constant false-alarm rate (CFAR) detector \cite{piofmd, 9399786}, and then perform subsequent processing only on the selected regions.

The CFAR-based target detection and localization algorithm operates on the raw radar I-Q samples vide \eqref{sifmimo} for each frame as presented in Fig.~\ref{cfar_target_det}(a). First, the Range and Velocity FFTs are performed on I-Q data in a frame to obtain the range-velocity (RV) map for initial target detection. The RV maps from all receivers are integrated non-coherently (i.e., sum the magnitude RV maps) to increase the signal to noise ratio of resulting RV map \cite{9107493}.

Post summing, the cell-averaging CFAR \cite{piofmd} algorithm is applied to detect targets and obtain their 2D (range and velocity) localization. During the CFAR detection process, each cell/bin is evaluated for the presence or absence of a target using a threshold, and the threshold adapts itself according to the noise power estimate within a sliding window.

Thereafter, peak grouping for all CFAR detected targets is done based on their 2D localization - see Fig.~\ref{cfar_target_det}(b) - by checking if the detected amplitude is greater than that of its neighbors. For illustration of Fig.~\ref{cfar_target_det}(b), in the $3\times3$ kernel centered at CFAR detection 1, CFAR detection 2 will be discarded if it is with smaller amplitude. At the end, we calculate the Angle FFT \footnote{Before this Angle FFT, we compensate the motion-induced phase error (shown in Section. \ref{angle_estimat}) for TDM-MIMO, using the detected Doppler velocity.} for each detected target across the RV maps of all receivers (in the virtual array formed by TDM-MIMO) to estimate its azimuth angle.

Assume a group of targets are detected from the first frame's radar data with above algorithm. For each detection with range and azimuth angle $(r, \theta)$, its relative Cartesian position $(x_{\R{r}}, y_{\R{r}})$ with respect to radar is given by $x_{\R{r}} = r\sin{\theta}, \, y_{\R{r}}  = r\cos{\theta}$.

\begin{figure}
  \centering
  \includegraphics[width=3.2in, trim=2 3 2 2,clip]{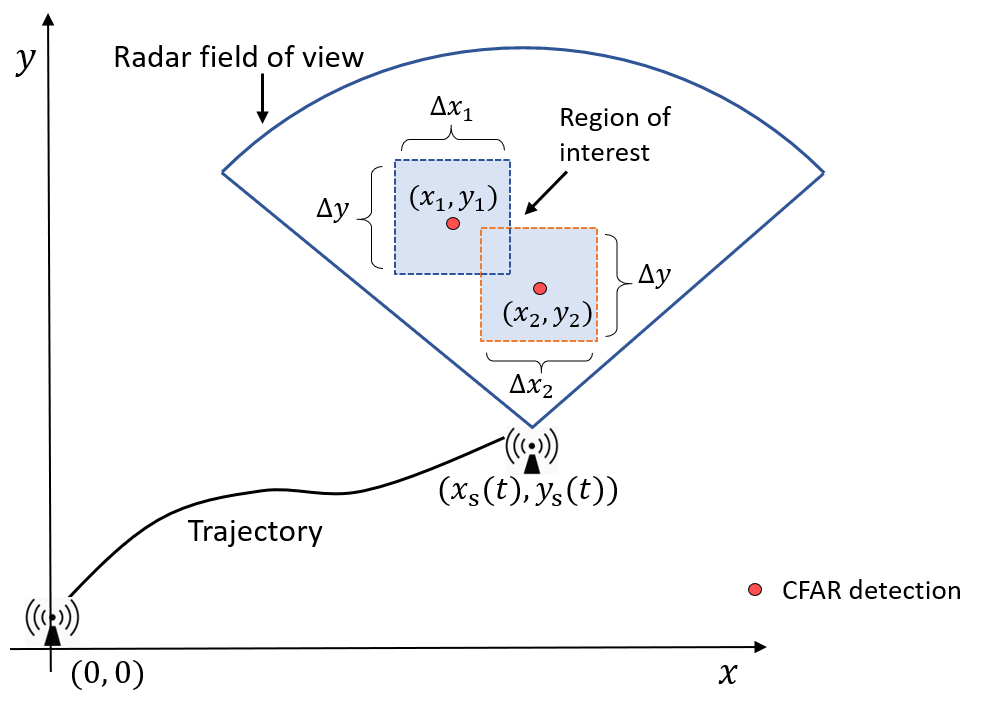}
  \caption{A example that shows how to define the ROI based on the localization of 2 detected targets $(x_1, y_1), \, (x_2, y_2)$.}
  \label{def_reg}
\end{figure}

The relative position $(x_{\R{r}}, y_{\R{r}})$ is transformed to the position $(x, y)$ in global geometry plane (see Fig.~\ref{def_reg}) by adding radar trajectory$(x_{\R{s}}(t), y_{\R{s}}(t))$, i.e., $(x,y) = (x_{\R{r}}, y_{\R{r}}) + (x_{\R{s}}(t), y_{\R{s}}(t))$. Based on the calculated global positions of targets, say $\{(x_1, y_1),(x_2, y_2)\}$ for example, we then can define the ROI $U_{\R{g}}$ in geometry plane that need to be imaged:
\begin{dmath}
     \Big\{ (x^\prime, y^\prime) \, | \, \{x_1 - \frac{\Delta x_1}{2} \le x^\prime \le x_1 + \frac{\Delta x_1}{2}, \, y_1 - \frac{\Delta y}{2} \le y^\prime \le y_1 + \frac{\Delta y}{2}\} \, \cup \, \{x_2 - \frac{\Delta x_2}{2} \le x^\prime \le x_2 + \frac{\Delta x_2}{2}, \, y_2 - \frac{\Delta y}{2} \le y^\prime \le y_2 + \frac{\Delta y}{2} \} \Big\}
\end{dmath}

\noindent where the overall regions are the union of two small regions that are centered on $(x_1, y_1), \, (x_2, y_2)$ respectively. The $\Delta x_1, \, \Delta x_2, \, \Delta y$ are the side length of two small regions. For each region, we use a fixed side length $\Delta y$ along y-axis, and customize the side length along x-axis for defined azimuth scope $\Delta \theta$ (i.e., $\Delta x = r \Delta \theta$).

The radar trajectory $(x_{\R{s}}(t), y_{\R{s}}(t))$ is required for defining ROI $U_{\R{g}}$ (indicated above), as well as for the phase compensation for coherent SAR processing. In Section \ref{rad odo}, we explain the radar odometry algorithm that takes the radial velocities and azimuth angles of detected targets as input to estimate the moving trajectory of ego-radar.

\subsection{Backprojection SAR Processing}

\begin{figure}
\centering
\includegraphics[width=3.45in, trim=1 3 1 1,clip]{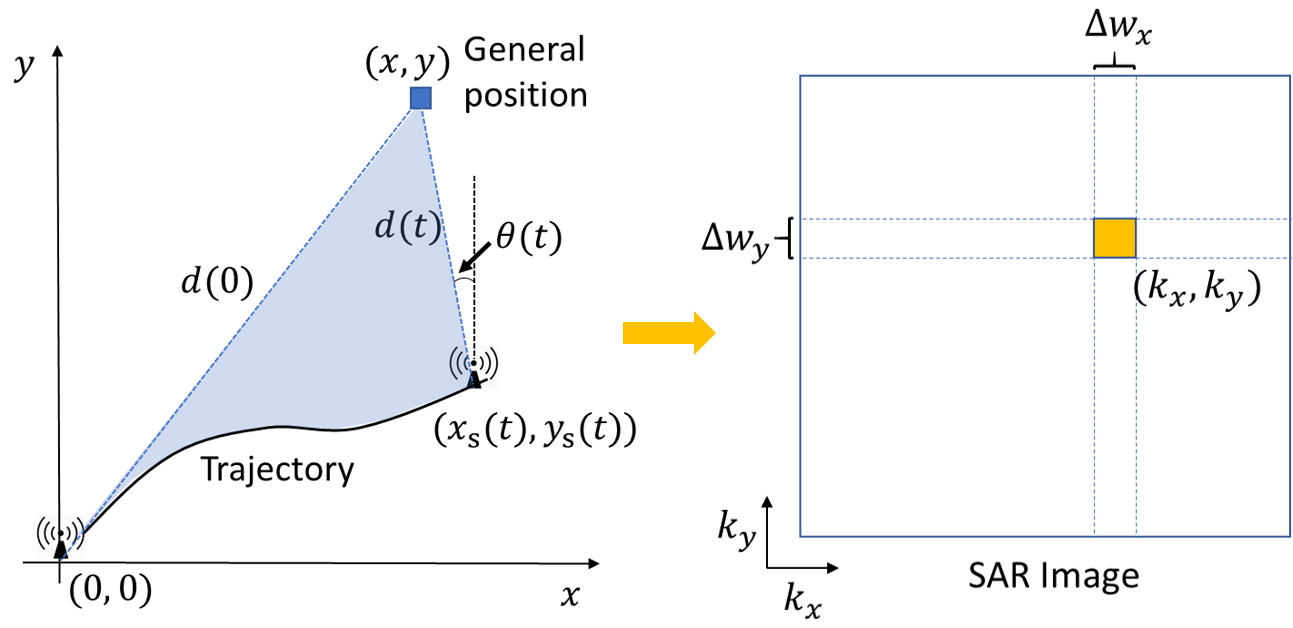}
\caption{The transformation from a general position $(x,y)$ in geometry plane to the pixel $(k_{\R{x}},k_{\R{y}})$ in SAR image plane.}
  \label{sarmap}
\end{figure}

In last section, we use the MIMO processing to define ROI $U_{\R{g}}$ in the geometry plane that need to be imaged. In this section, we would operate the backprojection algorithm on ROI to get the high-resolution image. Firstly, the region $U_{\R{g}}$ in geometry plane is transformed to a grid of pixels $U_{\R{s}}$ in the SAR image plane. As 
illustrated in Fig.~\ref{sarmap}, a general position $(x,y)$ in geometry plane corresponds to the pixel $(k_{\R{x}},k_{\R{y}})$ in the SAR image plane:
\begin{equation}
    x  = k_{\R{x}} \Delta w_{\R{x}}, \; y =  k_{\R{y}} \Delta w_{\R{y}}
    \label{mapp}
\end{equation}

\noindent where $\Delta w_{\R{x}}, \, \Delta w_{\R{y}}$ are the pixel width and height.

For each pixel $(k_{\R{x}},k_{\R{y}})$ in the $U_{\R{s}}$, we coherently sum up its corresponding radar measurement for multiple snapshots using matched filter response. Such algorithm updates the SAR image for every snapshot and reuses the previous calculations when updating with any new measurement. As presented in Fig.~\ref{system}, we select one snapshot from every $K$ chirps with 3D-FFT algorithm (Section \ref{sec_3dfft}) to reduce SAR sampling rates. The parameter $K$ is carefully chosen based on Section \ref{sec_prf} to avoid azimuth aliasing in defined scope $\Delta \theta$.

Using 3D-FFT \footnote{Note that we use small FFT points here to avoid the migration \cite{8695853} and also to reduce the computation complexity.}, the radar I-Q samples of $K$ chirps are converted to a RVA data cube where the corresponding radar measurement for pixel $(k_{\R{x}},k_{\R{y}})$ can be found. That is, for pixel $(k_{\R{x}},k_{\R{y}})$, we calculate its relative distance $d$ and azimuth angle $\theta$ to radar, and use them $(d, \theta)$ to localize the corresponding radar measurement in RVA data cube $S_{\R{RVA}}$ (i.e., a radar measurement is represented by a small red cube in Fig.~\ref{system}).

For example, we consider a train of $K$ chirps that start from $t$ (e.g., $t=nT_{\R{c}}+pT_{\R{f}}$ for chirp $n$ of frame $p$). At time $t$, the relative distance $d(t)$ and azimuth angle $\theta(t)$ for pixel $(k_{\R{x}},k_{\R{y}})$ is given by:
\begin{equation}
\begin{aligned}
     & d(t) = \sqrt{(x_{\R{s}}(t) - k_{\R{x}} \Delta w_{\R{x}})^2 + (y_{\R{s}}(t) - k_{\R{y}} \Delta w_{\R{y}})^2} \\
     & \theta(t) = \arcsin \left( \frac{x_{\R{s}}(t) - k_{\R{x}} \Delta w_{\R{x}}}{d(t)} \right)
\end{aligned}
  \label{disxy}
\end{equation}

The calculated $d(t)$ and $\theta(t)$ are mapped to the range bin $m_{\R{r}}$ and angle bin $m_{\theta}$ of the RVA cube respectively \cite{ramp} to localize the radar measurement: $m_{\R{r}} = \lfloor \frac{2M_{\R{r}} S_{\R{w}} d(t)}{c_0 f_{\R{s}}} \rfloor, \; m_{\theta} = \lfloor \frac{M_\theta h (x_{\R{s}}(t) - k_{\R{x}} \Delta w_{\R{x}})}{\lambda d(t)} \rfloor$. The velocity bin $m_{\R{v}}$ is chosen to have the max-intensity velocity component at the location $( m_{\R{r}} , m_{\theta})$. Therefore, the corresponding radar measurement for pixel $(k_{\R{x}},k_{\R{y}})$ at time $t$ is given by $S_{\R{RVA},t}(m_{\R{r}}, m_{\R{v}}, m_{\theta})$.

For coherent summation of the radar measurements from multiple snapshots, their phase must be compensated by a matched filter response $H(t)$ for the relative distance $d(t)$ \cite{glob} \cite{sar_tbx}. 
\begin{equation}
H(t)=\exp \left(j 2 \pi f_{\R{c}} \frac{2 d(t)}{c_0}\right)
\label{phasecom}
\end{equation}

The last step for backprojection SAR processing is to coherently sum the compensated radar measurements from all observed snapshots. For example, for $N_{\R{f}}$ frames as input where each frame contains $N_{\R{c}}/K$ snapshots, there are $N_{\R{f}}N_{\R{c}}/K$ snapshots in total. We denote the start time for the $l^{\R{th}}$ snapshot of frame $p$ by $t_{l,p}=(l-1)KT_{\R{c}}+pT_{\R{f}}$. Then the backprojection processing for pixel $(k_{\R{x}},k_{\R{y}})$ is given by:
\begin{dmath}
I\left(k_{\R{x}}, k_{\R{y}}\right)=\sum_{p=0}^{N_{\R{f}}} \sum_{l=0}^{N_{\R{c}}/K} H^{*}(t_{l,p})\cdot S_{\R{RVA},t_{l,p}}(m_{\R{r}}, m_{\R{v}}, m_{\theta})
\label{sarp}
\end{dmath}

To form an image using this method, we apply \eqref{sarp} for each pixel in ROI $U_{\R{s}}$, and leave the rest of the pixels in SAR image blank.

\section{Ego-motion Estimation with Radar Odometry \label{rad odo}}

As mentioned earlier, one benefit of FMCW MIMO radar is access to accurate estimate of target's radial Doppler velocity and azimuth angle within a frame \cite{gao2019experiments}. This in turn enables radar odometry since the sensor's velocity can be estimated by analyzing the relationship between the radial Doppler velocities and azimuth angles of all static targets in the field of view \cite{6728341, 9096265}. 

\begin{figure}
    \centering
    \includegraphics[width=3.1in, trim=2 3 2 2,clip]{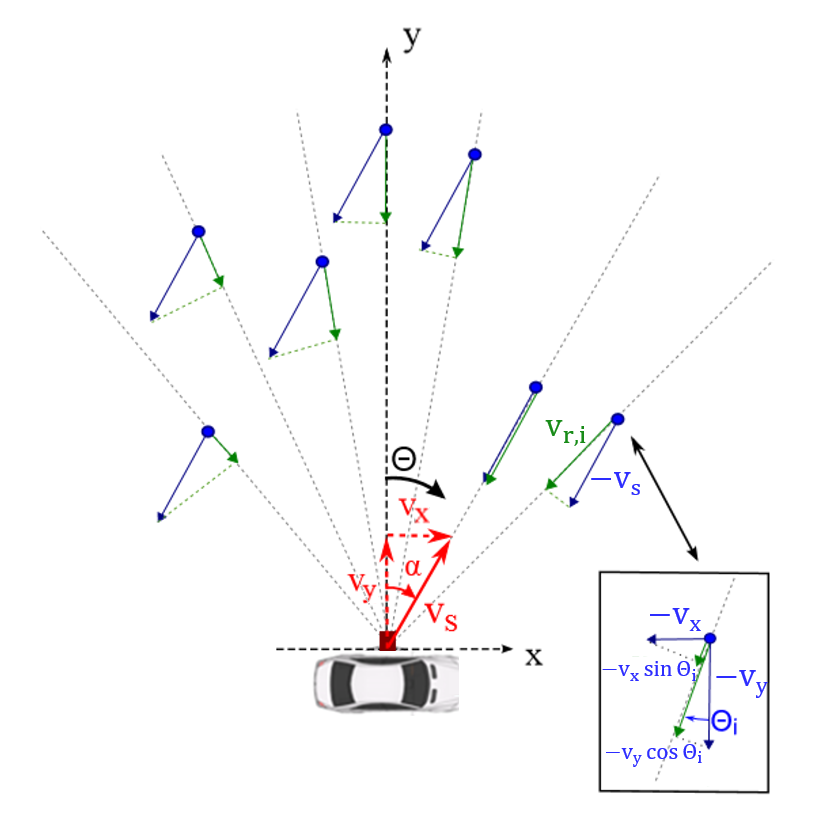}
    \caption{Illustration for radar odometry algorithm.}
  \label{odo}
\end{figure}

As illustrated in Fig.~\ref{odo}, if a radar sensor is moving with $v_{\R{s}}$, then all stationary targets move with relative velocity (blue line) equal to the sensor's speed with opposite heading. Given that Doppler radar only measures the radial velocity component (green line) of target, we can reconstruct the sensor's velocity components along x-axis and y-axis $(v_{\R{x}}, v_{\R{y}})$ by analyzing the velocity profile of at least two stationary targets \cite{6728341, 9096265}. 

For $N_{\R{d}}$ detected stationary targets in a frame, the velocity profile of these targets is given by \eqref{odo_equ}, where $v_{\R{r}, i}$ and $\theta_{i}$ are the measured radial Doppler velocity and azimuth angle for $i^{\R{th}}$ stationary target \cite{6728341}.
\begin{equation}
    \left[\begin{array}{c}
    v_{\R{r}, 1} \\
    \vdots \\
    v_{\R{r}, N_{\R{d}}}
    \end{array}\right]=-\left[\begin{array}{cc}
    \sin \left(\theta_{1}\right) & \cos \left(\theta_{1}\right) \\
    \vdots & \vdots \\
    \sin \big(\theta_{N_{\R{d}}}\big) & \cos \big(\theta_{N_{\R{d}}}\big)
    \end{array}\right]\left[\begin{array}{c}
    v_{\R{x}} \\
    v_{\R{y}}
    \end{array}\right]
    \label{odo_equ}
\end{equation}


We solve \eqref{odo_equ} via the minimum mean-square-error (MMSE) estimator \cite{wiki:mmse} to obtain the frame-level radar velocity $\bm{\hat{v}_{\R{s}}}=(\hat{v}_{\R{x}}, \hat{v}_{\R{y}})$. We denote the estimated sensor velocity for frame $p$ by $\bm{\hat{v}}_{\R{s},p} = (\hat{v}_{\R{x},p}, \hat{v}_{\R{y},p})$ and assume that radar moves with constant velocity within frame \cite{8995552}. Then the position of radar $(x_{\R{s}}(t), \, y_{\R{s}}(t))$ at time $t$ is obtained by integrating the estimated sensor velocity profile. Considering the chirp $n$ of frame $p$ that starts at time $t=nT_{\R{c}}+pT_{\R{f}}$, the corresponding radar position estimation $(\hat{x}_{\R{s}}(t), \, \hat{y}_{\R{s}}(t))$ is given by:
\begin{equation}
\begin{aligned}
     & \hat{x}_{\R{s}}(t) = \sum_{u=0}^{p-1} \hat{v}_{\R{x},u} T_{\R{f}} + \hat{v}_{\R{x},p} n T_{\R{c}} \\
     & \hat{y}_{\R{s}}(t) = \sum_{u=0}^{p-1} \hat{v}_{\R{y},u} T_{\R{f}} + \hat{v}_{\R{y},p} n T_{\R{c}}
\end{aligned}
\label{linermotion}
\end{equation}

\noindent where the initial location of radar is set to the origin without loss of generality, $T_{\R{f}}$ is the frame duration, and $T_{\R{c}}$ is the chirp duration. 

In reality, moving objects are expected in an experimental scenario, resulting in mixing detection of stationary objects and moving objects. Therefore, we employ Random Sample Consensus (RANSAC) algorithm \cite{10.1145/358669.358692} prior to radar odometry to separate out needed stationary targets and determine $N_{\R{d}}$ \cite{6728341}. RANSAC is an iterative method for optimal extraction of inliers (corresponding to stationary targets) that fit model \eqref{odo_equ} very well and separation of outliers (moving targets or clutter) by random sampling of observed data \cite{10.1145/358669.358692}.


\begin{table*}[h]
\begin{center}
\caption{system parameter calculation (based on \cite{gao2019experiments}) and configurations for mimo-sar processing}
\begin{tabular}{ll|ll}
\toprule  
Parameter & Calculation Equation& Configuration & Value\\
\midrule  
Range resolution ($R_{\R{res}}$) & $R_{\R{res}}=\frac{c_0}{2B}=\SI{0.447}{m}$  &  Frequency ($f_{\R{c}}$) & \SI{77}{GHz}\\[0.75ex]

Velocity resolution ($V_{\R{res}}$) &  $V_{\R{res}}=\frac{\lambda}{2N_{\R{c}} T_{\R{c}}}=\SI{0.0848}{m/s}$ & Sweep Bandwidth ($B$) & \SI{335}{MHz}\\[0.75ex]

Angle resolution ($\theta_{\R{res}}$)  &  $\theta_{\R{res}}=\frac{\lambda}{N_{\R{Tx}}N_{\R{Rx}}h\cos{\theta}} \approx \ang{15}$  & Sweep slope ($S_{\R{w}}$) & \SI{21}{MHz \per\micro\second} \\[0.75ex]

Max operating range ($R_{\R{max}}$) & $R_{\R{max}} = \frac{f_{\R{s}} c_0}{2S_{\R{w}}} = \SI{28.5}{m}$ & Sampling frequency ($f_{\R{s}}$) & \SI{4000}{ksps}\\[0.75ex]

Max operating velocity ($V_{\R{max}}$) & $V_{\R{max}} = \frac{\lambda}{4T_{\R{c}}} = \SI{10.82}{m/s}$ & Num of chirps in one frame ($N_{\R{c}}$) & $255$ \\[0.75ex]

Max operating angle ($\theta_{\R{max}}$) & $\theta_{\R{max}} = \sin^{-1}{\left(\frac{\lambda}{2h}\right)} = 90\degree$ & Num of samples of one chirp ($N_{\R{s}}$) & $64$ \\[0.75ex]

 & & Num of transmitters, receivers ($N_{\R{Tx}}$, $N_{\R{Rx}}$) & $2$, $4$\\[0.75ex]
 
 & & Duration of chirp~\protect \footnotemark~and frame ($T_{\R{c}}$, $T_{\R{f}}$) & \SI{90}{\micro\second}, \SI{33.3}{ms}\\
\bottomrule 
\label{param}
\end{tabular}
\end{center}
\end{table*}

\footnotetext{$T_{\R{c}}$ is equal to chirp interval times number of Tx.}

\section{System Parameter Requirement \label{sys para requr}}

\subsection{Pulse Repetition Frequency (PRF) \label{sec_prf}}

The along-track sampling/update rate is a significant limitation of SAR imaging. Here, we use the PRF to represent the sampling rate. The PRF must be greater than the azimuth signal bandwidth to avoid azimuth resolution ambiguities caused by aliasing \cite{Cumming2005DigitalPO}. By the Nyquist sampling theorem, it is usual to ensure that the SAR element spacing $\Delta u$ is less than $\lambda / 2$ to avoid the aliasing lobes \cite{126956, 5960684}. For the moving radar with velocity $v_{\R{s}}$, wavelength $\lambda$, PRF $f_{\R{p}}$, the azimuth sampling constraint can be expressed as: $v_{\R{s}} \cdot \frac{1}{f_{\R{p}}} < \frac{\lambda}{2}$. This requires either increasing the PRF (reduce $K$ in Fig.~\ref{system}), or limiting the maximum velocity of source radar. 

For the proposed MIMO-SAR algorithm, the above azimuth sampling requirement might be relaxed by considering the potential aliases only within a small region (as shown in Fig.~\ref{def_reg}). From \cite{126956}, the minimum required sampling space $\Delta u$ for imaging the defined small region that has azimuth scope $\Delta \theta$ without alias is given by $\Delta u < \lambda / \left( 2\sin{\frac{\Delta \theta}{2}} \right)$. Hence, we can obtain the sampling constraint for moving radar with MIMO-SAR:
\begin{equation}
  v_{\R{s}} \cdot \frac{1}{f_{\R{p}}} < \frac{\lambda}{2\sin{\frac{\Delta \theta}{2}}}
  \label{prf}
\end{equation}

For example, the along track sampling for \SI{77}{GHz} radar should be finer than \SI{2}{mm}. Given $f_{\R{p}} = \SI{556}{Hz}$ (SAR updating every \SI{1.8}{ms}, $K=20$ in Fig.~\ref{system}), radar must travel no faster than \SI{1.1}{m/s}, which is unrealistic for automotive scenario. Limiting to the azimuth scope $\Delta \theta=5\degree$ for each small region, radar should move no faster than \SI{25.1}{m/s}. In practice, the chosen PRF is little higher than the minimum requirement for oversampling.

\subsection{Coherent Processing Interval (CPI)}
The CPI for SAR processing is defined as the total duration over which the receive echo can be coherently processed. We assume that the coherent phase course can no longer be guaranteed when the matched filter response \eqref{phasecom} has a phase error greater than threshold $\pi / 2$, resulting in loss of image quality. We analyze two sources of the phase error in \eqref{phasecom}: system clock error and velocity estimation error from ego-motion. 


The matched filter response is $\exp \left(j 2 \pi f_{\R{c}} \frac{2 d}{c_0}\right)$, where the inaccurate carrier frequency $f_{\R{c}}$ and distance $d$ can result in the phase error. If the system clock provided by local oscillator has error, there will be a deviation on both $f_{\R{c}}$ and the measurement timing (therefore affecting $d$). According to \eqref{disxy} \eqref{linermotion}, any velocity error in estimating radar's ego motion will also affect $d$. Based on \cite{ti1843datasheet}, the maximum clock error for TI's on-board \SI{40}{MHz} oscillator is \SI{200}{ppm}, which is very small compared to the potential velocity error. We therefore only consider the affect of velocity error on CPI in the following analysis.

We assume the velocity errors $\epsilon_{v_{\R{x}}}, \, \epsilon_{v_{\R{y}}}$ for any frame are independent Gaussian variables, i.e., $\epsilon_{v_{\R{x}}} \sim \mathcal{N} (0, \sigma^{2}), \, \epsilon_{v_{\R{y}}}  \sim \mathcal{N} (0, \sigma^{2})$. Then the distance deviation $\Delta d$ at frame $p$ is the accumulation of the previous velocity error components projected to the radial direction $\theta_i$. 
\begin{equation}
\Delta d_p = \abs{ \sum_{i=0}^{p-1} \left( \cos{\theta_i} \cdot \epsilon_{v_{\R{x}}, i} + \sin{\theta_i} \cdot \epsilon_{v_{\R{y}}, i} \right) } \cdot T_{\R{f}}
\end{equation}

Since $\epsilon_{v_{\R{x}}, i}, \epsilon_{v_{\R{y}}, i}$ are all independent Gaussian variables, we have:
\begin{equation}
\begin{aligned}
    \cos{\theta_i } \cdot \epsilon_{v_{\R{x}}, i}  + \sin{\theta_i} \cdot \epsilon_{v_{\R{y}}, i} \sim \mathcal{N} (0, \sigma^{2} ) \\
    \sum_{i=0}^{p-1} \left( \cos{\theta_i} \cdot \epsilon_{v_{\R{x}}, i} + \sin{\theta_i} \cdot \epsilon_{v_{\R{y}}, i} \right) \sim \mathcal{N} (0, p \sigma^{2})
\end{aligned}
\end{equation}

Therefore, $\Delta d_p$ follows the folded normal distribution \cite{wiki:xxx} with expectation $\E \left[ \Delta d_p \right] = \sqrt{\frac{2 p}{\pi}} \sigma T_{\R{f}}$. Based on \eqref{phasecom}, the relative phase error caused by $\E \left[ \Delta d_p \right]$ can be expressed as $\Delta \phi_p = 4\pi f_{\R{c}} \E \left[ \Delta d_p \right] / c_0$. When $\Delta \phi_p$ exceeds a threshold $\phi_{\R{Th}}$, we define the limits of CPI. Thus, the number of frames $N_{\R{f}}$ within CPI is given by: 
\begin{equation}
N_{\R{f}}=\frac{1}{2 \pi}\left(\frac{c_0 \cdot \phi_{\R{Th}}}{4 f_{\R{c}} \sigma T_{\R{f}}} \right)^{2}
\label{NF}
\end{equation}

For example, given $\phi_{\R{Th}}=\pi / 2$, $f_{\R{c}}= \SI{77}{GHz}$, $T_{\R{f}}=\SI{33.3}{ms}$, $\sigma=0.005$, the CPI $N_{\R{f}} \approx 14$ frames.

\subsection{Region of Interest (ROI)}
The definition of ROI in Section~\ref{MIMO SAR algo} is important for balancing computational efficiency and overall imaging quality. Each ROI square is defined by two parameters: side length along y-axis $\Delta y$ and azimuth scope $\Delta \theta$. With smaller $\Delta y$ and $\Delta \theta$, the ROI covers smaller area for imaging, thus resulting in lower computation load. Besides, Equation \eqref{prf} implies that smaller $\Delta \theta$ gives lower PRF requirement as well for avoiding azimuth aliasing in ROI. However, it is possible to miss interested targets in imaging results if we choose too small $\Delta y$ or $\Delta \theta$, i.e., when the resulting ROI is not big enough to cover the undetected targets of CFAR algorithm. In following experiments, we use a sensitive CFAR detector with probability of false alarm $10^{-4}$, and select medium-size ROI with $\Delta y=\SI{0.9}{m}$ and $\Delta \theta=\SI{5}{\degree}$ accordingly to achieve necessary balance.

\section{Simulations \label{simu res}}

The reported simulations were conducted using MATLAB R2019b on a computer with Intel i7-9750H CPU. In particular, we utilize the MATLAB Automated Driving Toolbox \cite{Matlabtool} to simulate the driving scenario in Section \ref{sec_simu_parlot} and \ref{sec_simu_roadside} for MIMO-SAR imaging.

\subsection{Imaging Point Targets \label{imag_pot}}

\subsubsection{Simulation Studies via MATLAB}
Assume two static point targets in the field of view at same range (\SI{5}{m}) but at angles $\left(\ang{-0.5} \text{ and } \ang{0.5} \right)$ that are within a current angular beam-width (hence cannot be separated by conventional 3D-FFT processing). 
The initial location of the radar sensor is $\left(0,0\right)$, and it is assumed to move with velocity $\left(v_{\R{x}}, v_{\R{y}}\right) = \left(\SI{1}{m/s}, 0\right)$. Other system configurations for radar are presented in Table \ref{param}. We use MATLAB to simulate the TDM-MIMO radar with the I-Q samples vide \eqref{sifmimo} post-demodulated at the receiver.

\subsubsection{Range-Angle Imaging for MIMO Radar \label{range-angle imaging}}

To compare the imaging capability of different algorithms, we show the range-angle image \cite{gao2019experiments} for MIMO radar data. The range-angle maps are obtained by operating Range and Angle FFTs on the simulated I-Q samples of first frame. We then average the amplitude range-angle maps of all chirps to get the final image.

\begin{figure}
    \centering
    \includegraphics[width=3in, trim=2 5 2 2,clip]{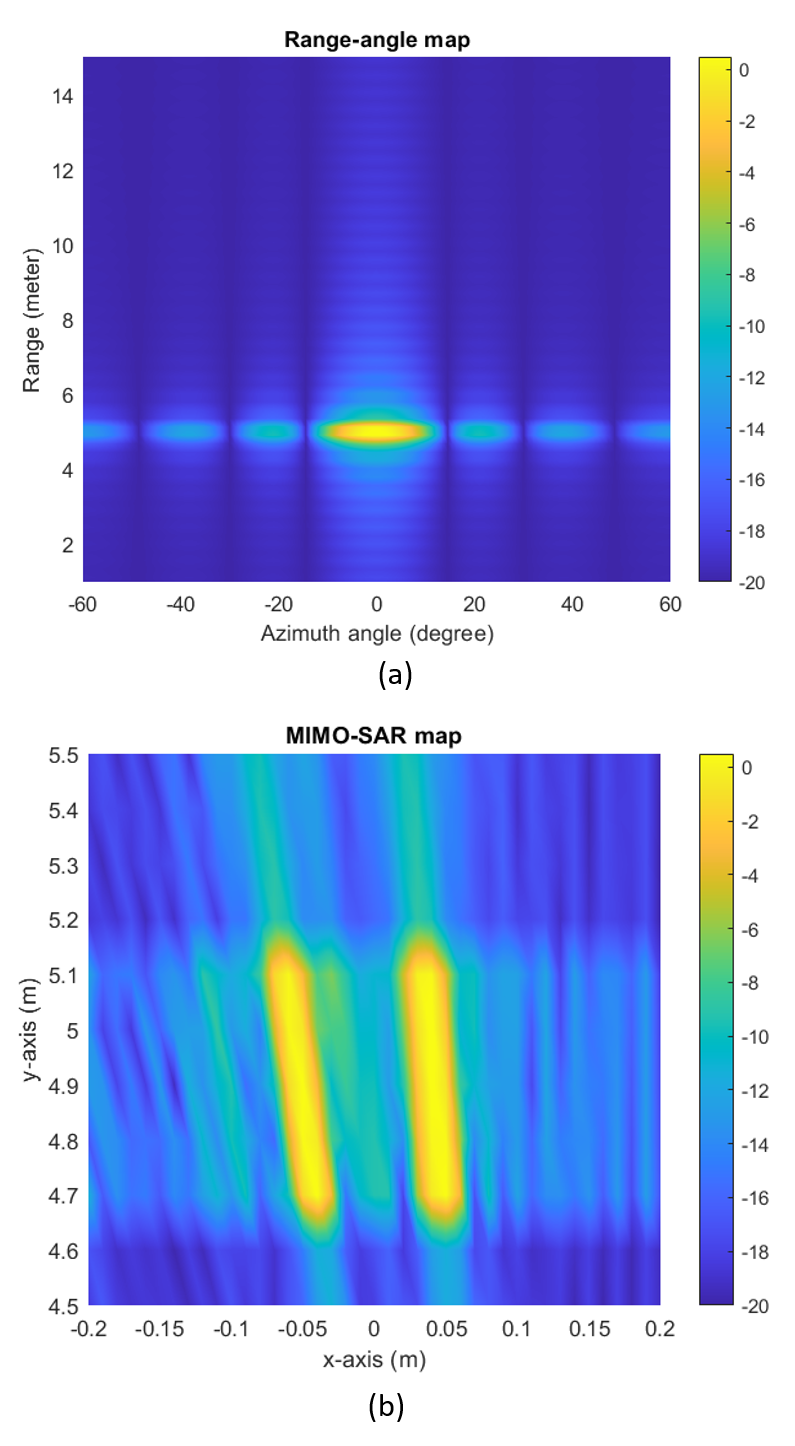}
    \caption{Imaging two close point targets: (a) Range-angle map imaging for the fist frame; (b) MIMO-SAR imaging (magnitude $I$ of \eqref{sarp}) for 13 frames.}
  \label{point_targ_imags}
\end{figure}

As shown in Fig.~\ref{point_targ_imags}(a), the generated range-angle map presents the imaging for two point targets with very close azimuth angles. Results tell that two point targets can not be separated with current MIMO beam-width (around \ang{15} resolution).

\subsubsection{MIMO-SAR Imaging}
Using the MIMO operation (CFAR detection and peak grouping), we select the ROI that covers two point targets. The selected region covers the \SI{-0.2}{m} $\sim$ \SI{0.2}{m} (\ang{4.6} coverage) in x-axis and \SI{4.5}{m} $\sim$ \SI{5.5}{m} in y-axis. The spacing of x-axis and y-axis are \SI{0.01}{m} and \SI{0.1}{m} respectively. We assume that radar velocity is known at the receiver in this simulation, and perform the backprojection SAR processing on each pixel of ROI for 13 frames. The parameters for MIMO-SAR are set according to Table \ref{para_sar}. 

As shown in Fig.~\ref{point_targ_imags}(b), the resulting MIMO-SAR image shows a clear separation between two nearby point targets in azimuth, thereby validating that the MIMO-SAR imaging algorithm can effectively increase azimuth angle resolution, and achieve separation of close point targets.

\begin{table}
\begin{center}
\caption{parameter values for mimo-sar algorithm}
\begin{tabular}{ll}
\toprule  
Parameter & Value \\
\midrule  
Num of chirps for SAR processing ($K$) & $20$\\
Pixel size of SAR image ($k_{\R{x}}$, $k_{\R{y}}$) & \SI{0.01}{m}, \SI{0.1}{m} \\
y-axis side length for each ROI ($\Delta y$) & \SI{0.9}{m}\\
Azimuth scope for each ROI ($\Delta \theta$) & \ang{5} \\
Range FFT points & $64$ \\
Velocity FFT points for target detection & $256$ \\
Velocity FFT points for SAR processing & $20$ \\
Angle FFT points for target localization & $128$ \\
Angle FFT points for SAR processing & 16 \\
\bottomrule 
\vspace{-3mm}
\label{para_sar}
\end{tabular}
\end{center}
\end{table}

\subsection{Verifying Predicted CPI for MIMO-SAR}

From \eqref{NF}, the CPI for SAR processing is mainly determined by the root-mean-square velocity error $\sigma$ of radar's ego-motion. With increasing time, the accumulated velocity error causes an increase in phase error in the matched filter response \eqref{phasecom}, thus limiting the CPI. To validate the predicted CPI, we reuse the simulation environment in Section \ref{imag_pot}, add r.m.s. ego-motion velocity error $\sigma$, and evaluate the phase error of a pixel in ROI versus time. This simulation was averaged 1000 times to plot the mean phase error vs. time curves for different velocity error values.

\begin{figure}[h]
    \centering
    \includegraphics[width=3.4in, trim=2 3 2 2,clip]{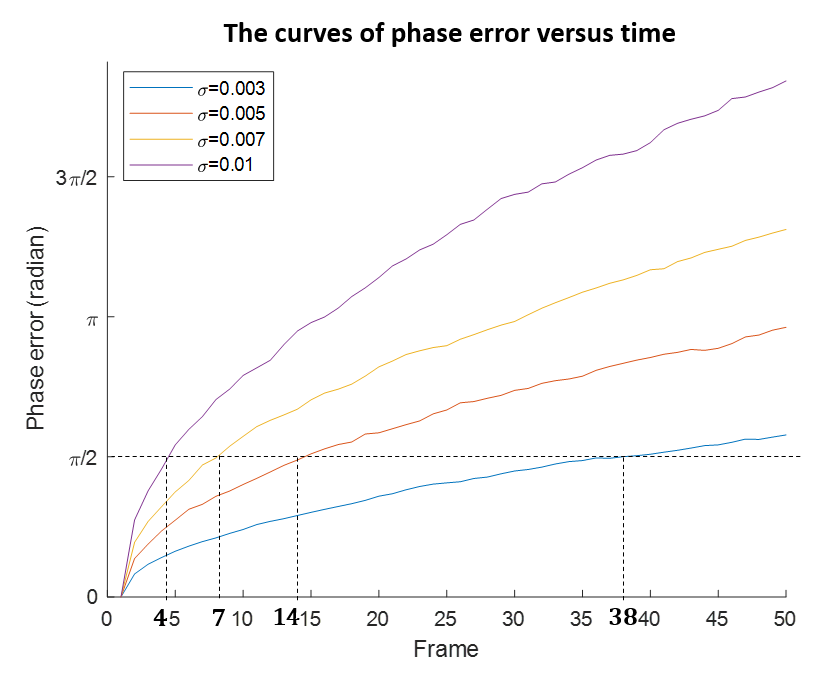}
    \caption{The curves of phase error versus time, for velocity error $\sigma = 0.003, \, 0.005, \, 0.007, \, 0.01$.}
  \label{phase_time_plot}
\end{figure}

We show the curves for velocity error $0.003$, $0.005$, $0.007$, and $0.01$ respectively in Fig.~\ref{phase_time_plot}, and mark the frame points that attain the phase error threshold $\pi / 2$. For example, given $\sigma = 0.003$, the calculated CPI from \eqref{NF} is $38$ frames. It shows that in the blue curve of Fig.~\ref{phase_time_plot}, phase error of frame $38$ is nearest to $\pi / 2$. Similarly, we validate that the CPIs for velocity error $0.005, \, 0.007, \, 0.01$ are $14, \, 7, \, 4$ frames respectively.

\begin{figure*}
    \centering
    \includegraphics[width=7.0in, trim=1 3 1 1,clip]{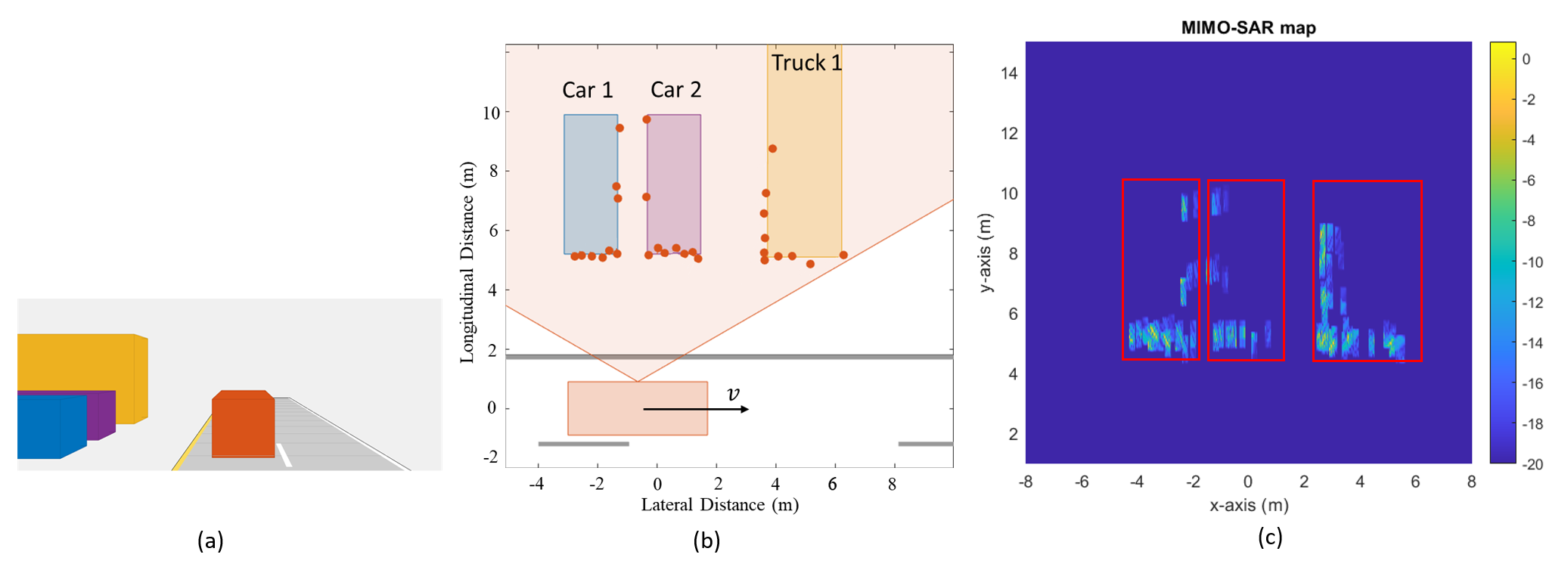}
    \caption{MIMO-SAR simulation for the parking lot scenario: (a) The ego-centric view of the environment with one driving car, two parked cars and 1 truck; (b) The bird-view for the parking lot environment; The red points are the point-representation of the extended cars and truck; (c) MIMO-SAR imaging result (magnitude $I$ of \eqref{sarp}) for parking lot scenario where we use three rectangles to cover the imaging cars and truck.}
  \label{parking_lot_simula}
\end{figure*}

\begin{figure*}
    \centering
    \includegraphics[width=7.0in, trim=1 3 1 1,clip]{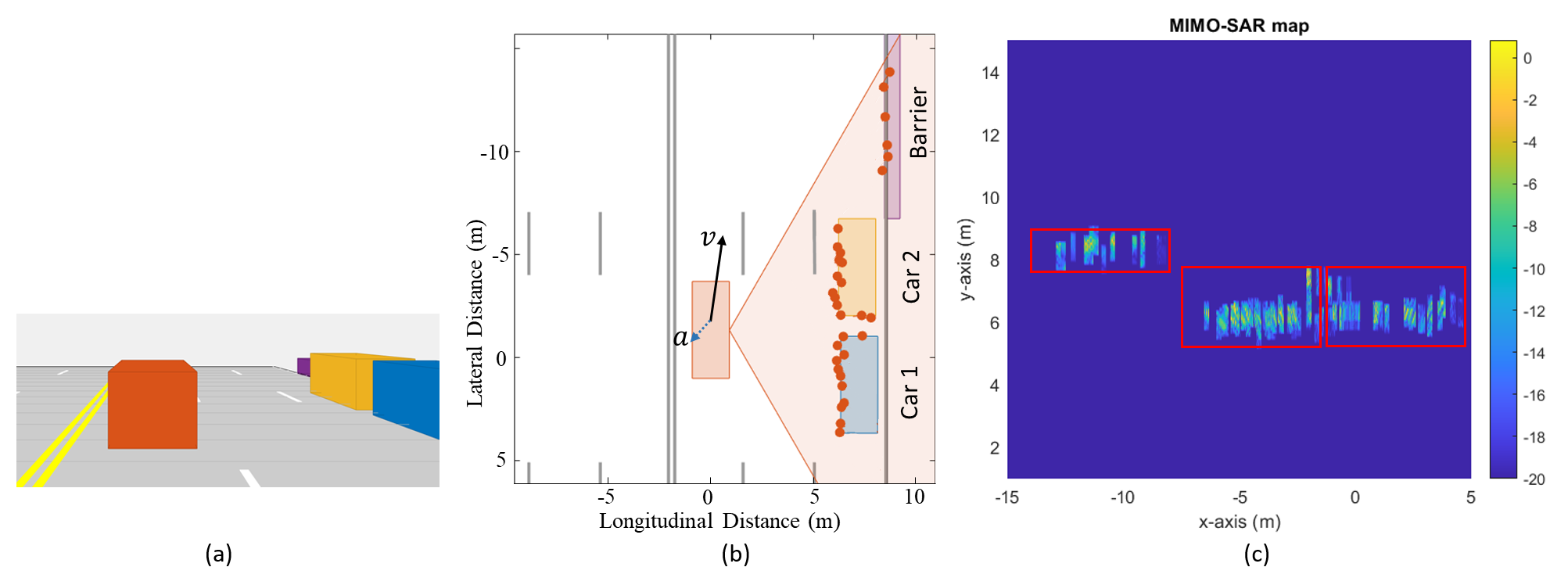}
    \caption{MIMO-SAR imaging simulation for the roadside scenario: (a) The ego-centric view of the environment with one driving car, two parked cars and 1 roadside barrier; (b) The bird-view for the roadside environment; The red points are the point-representation of the extended cars and barrier; (c) MIMO-SAR imaging result (magnitude $I$ of \eqref{sarp}) for the roadside scenario where we use three rectangles to cover the imaging cars and barrier.}
  \label{road_side_simula}
\end{figure*}

\subsection{MIMO-SAR Imaging for Simulated Parking Lot Scenario \label{sec_simu_parlot}}
\subsubsection*{Simulation Environment}
We simulate the driving environment for parking lot with the MATLAB Automated Driving Toolbox \cite{Matlabtool} to validate the MIMO-SAR algorithm. As shown in Fig.~\ref{parking_lot_simula}(a), this simulation environment contains 1 (self) driving car, 2 parked cars, and 1 parked truck. The driving car is with constant moving velocity \SI{4}{m/s}, and integrated with a radar sensor on its left side to observe the parked vehicles. Each extended vehicle is modeled as a group of point targets (see Fig.~\ref{parking_lot_simula}(b)) in radar's field of view. We then simulate the radar I-Q samples for these point targets with parameters configured as Table. \ref{param}.

\subsubsection*{MIMO-SAR Imaging}
We perform the MIMO-SAR algorithm on simulated I-Q samples assuming that the moving velocity of radar is unknown. That is, before operating the backprojection SAR, we use the radar odometry to estimate the radar trajectory. The parameters for MIMO-SAR are set according to Table. \ref{para_sar}. 

The MIMO-SAR imaging result for the data from 3 frames is shown in Fig.~\ref{parking_lot_simula}(c) where we use three rectangles to highlight the parked cars and truck. The synthetic aperture for 3 frames is around \SI{0.4}{m} ($\SI{4}{m/s} \times 3 \times \SI{33.3}{ms} = \SI{0.4}{m}$), which is much larger than the intrinsic MIMO aperture ($4 \lambda \approx \SI{16}{mm}$). It turns out that MIMO-SAR is effective for imaging the boundary of extended objects and separating close objects with large synthetic aperture.

\begin{figure*}[t]
    \centering
    \includegraphics[width=7.3in, trim=1 3 1 1,clip]{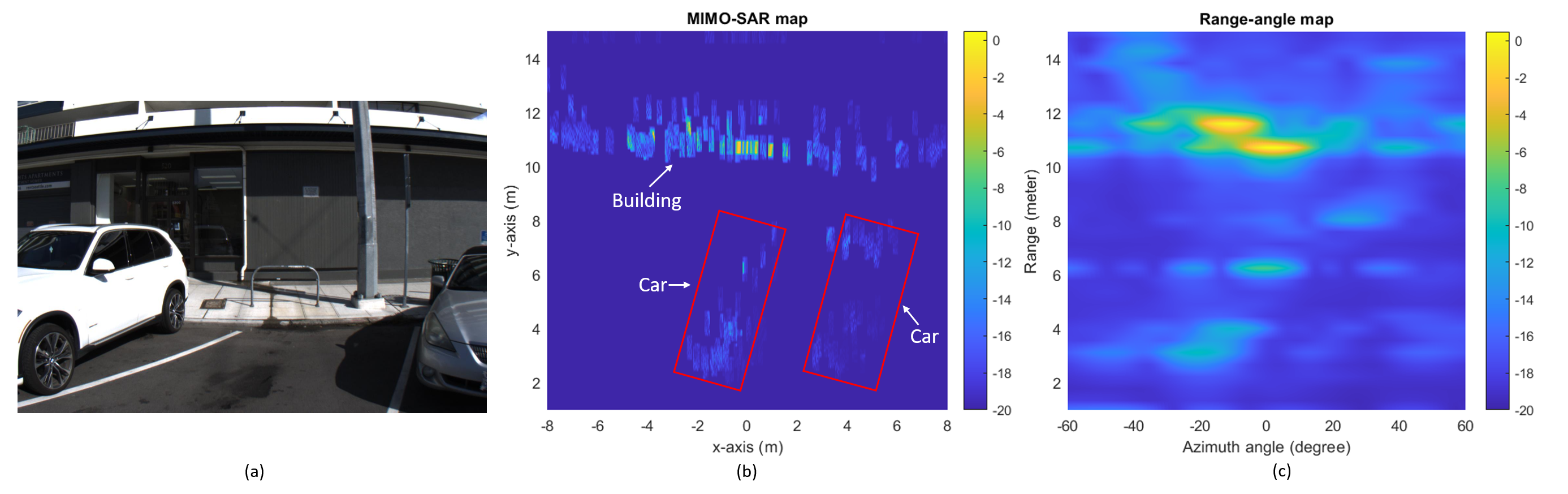}
    \caption{MIMO-SAR imaging for roadside experiment 1: (a) The camera image for the imaging environment with two inclinedly parked cars; (b) The MIMO-SAR imaging result (magnitude $I$ of \eqref{sarp}) where we use two rectangles to cover the parked cars; (c) Range-angle map imaging for single-frame radar data.}
  \label{road_side_exper2}
\end{figure*}

\begin{figure*}[t]
    \centering
    \includegraphics[width=7.3in, trim=1 3 1 1,clip]{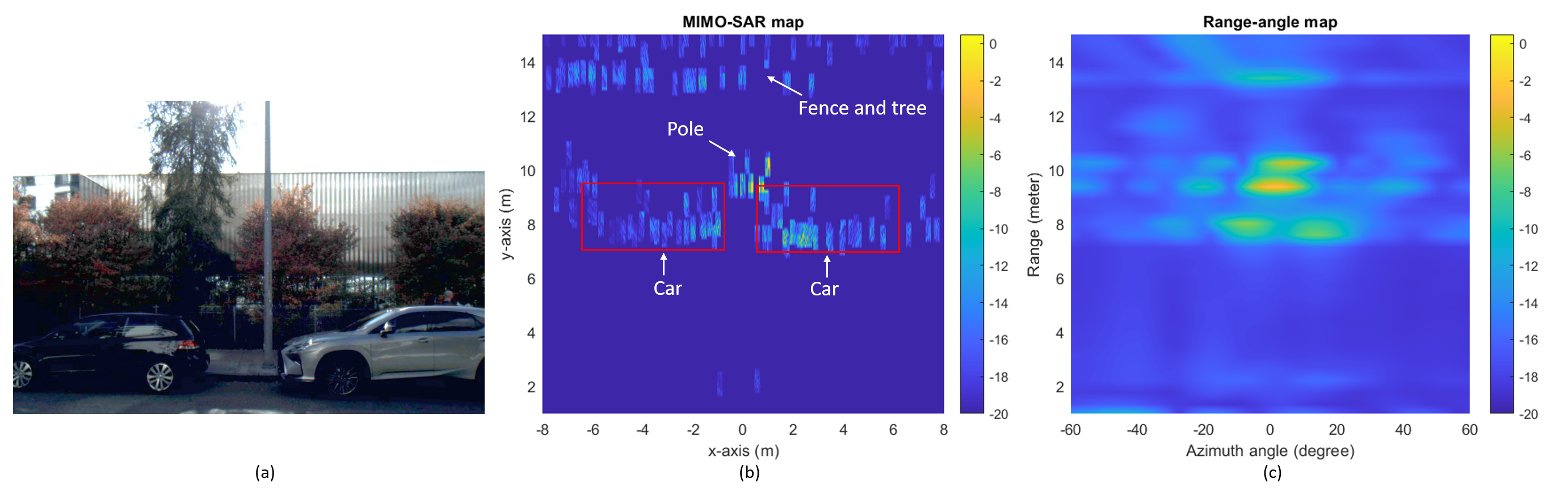}
    \caption{MIMO-SAR imaging for roadside experiment 2: (a) The camera image for the imaging environment with two parallelly parked cars; (b) MIMO-SAR imaging result (magnitude $I$ of \eqref{sarp}) where we use two rectangles to cover the parked cars; (c) Range-angle map imaging for single-frame radar data.}
  \label{road_side_exper3}
\end{figure*}

\subsection{MIMO-SAR Imaging for Simulated Roadside Scenario \label{sec_simu_roadside}}
\subsubsection*{Simulation Environment}

We simulate the new roadside scenario with MATLAB Automated Driving Toolbox \cite{Matlabtool}. As shown in Fig.~\ref{road_side_simula}(a), this simulation environment contains 1 (self) driving car, 2 parallelly parked cars, and 1 roadside barrier. The driving car is moving forward with initial velocity $(\SI{-6}{m/s}, \SI{1}{m/s})$ and acceleration $(\SI{2}{m/s^2}, \SI{-2}{m/s^2})$, and also integrated with a radar on its right side. Similar to above, the extended vehicles and barrier are modeled as a group of point targets within radar's field of view (see Fig.~\ref{road_side_simula}(b)) to simulate the post-modulated I-Q samples.

\subsubsection*{MIMO-SAR Imaging}
The MIMO-SAR algorithm accompanied with radar odometry is performed on the simulated I-Q samples to image above roadside environment. We show the MIMO-SAR map for 3 frames in Fig.~\ref{road_side_simula}(c) and use three rectangles to cover the imaging of parked cars and barrier.

\section{Experimental Results \label{exper res}}

\subsection{Experiment Platform Setup}

A set of camera images along with corresponding radar raw data (I-Q samples post demodulated at the receiver) \cite{ramp} have been collected by a vehicle-mounted platform (see Fig.~\ref{platform}(a)). The platform is placed on one side to facilitate sideways data collection whereby roadside objects are illuminated.

The data collection platform - see Fig.~\ref{platform}(b) - consists of 2 FLIR cameras (left and right) and a TI AWR1843 EVM radar \cite{ti1843evm}. As shown in Fig.~\ref{platform}(c), the AWR1843 radar is integrated with 2 Tx and 4 Rx \footnote{The $3^{\R{rd}}$ elevation transmitter on AWR1843 radar board is not used.}, which forms a 8-element horizontal virtual array with TDM-MIMO. We configure the parameters of this radar according to Table. \ref{param}. The binocular cameras are synchronized with radar to provide the visualization for imaging scenario. 
Since there is no any external navigation sensor (IMU or GPS) on the platform, we include radar odometry with RANSAC for all experiments below to estimate the necessary radar motion trajectory.



\begin{figure*}
    \centering
    \includegraphics[width=7.3in, trim=1 3 1 1,clip]{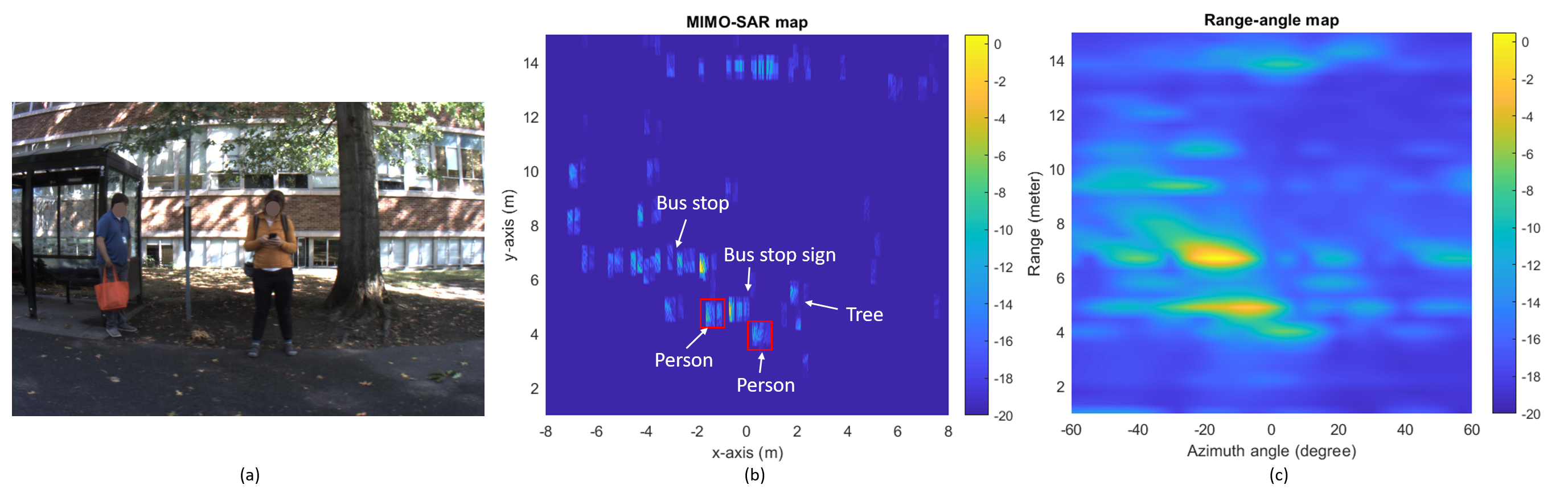}
    \caption{MIMO-SAR imaging for roadside experiment 3: (a) The camera image for the imaging environment with two static people; (b) MIMO-SAR imaging result (magnitude $I$ of \eqref{sarp}) where we use two rectangles to cover the standing persons; (c) Range-angle map imaging for single-frame radar data.}
  \label{road_side_exper4}
\end{figure*}

\subsection{Experiments}

For results using our own collected data, the MIMO-SAR algorithm parameter values used are listed in Table \ref{para_sar}; all post-processing computations were conducted using MATLAB R2019b on a computer equipped with Intel i7-9750H CPU.

\subsubsection*{\textbf{Roadside Experiment 1}}
As shown in Fig.~\ref{road_side_exper2}(a), the scene contains 2 parked cars roadside at an angle and a building in background. The vehicle-mounted platform moves with unknown velocity and trajectory (i.e., not collected by navigation sensor). We perform the MIMO-SAR algorithm on the collected radar I-Q samples to obtain the high-resolution imaging of this environment. The imaging result using 3 frames is presented in Fig.~\ref{road_side_exper2}(b), where the background building has well-defined boundary and two parked cars are clearly visible.
\vspace{0.4em}

\subsubsection*{\textbf{Roadside Experiment 2}}
The scene for second experiment contains 2 parallel parked cars and the background (fence and tree), as shown in Fig.~\ref{road_side_exper3}(a). Similar to above, the MIMO-SAR algorithm is operated on the collected radar I-Q samples lasting for 3 frames. We present the MIMO-SAR map in Fig.~\ref{road_side_exper3}(b) where the pole and two closely parked cars are well-separated.\vspace{0.4em}

\subsubsection*{\textbf{Roadside Experiment 3}}
The third experiment shown in Fig.~\ref{road_side_exper4}(a) is going to image 2 standing persons and background (bus stop and tree). Its MIMO-SAR imaging result from 3 frames is presented in Fig.~\ref{road_side_exper4}(b), from which we can identify the persons from the strong-reflection background (i.e., bus stop and the sign between two persons).

The above experiments validate the effectiveness of MIMO-SAR algorithm for imaging closely located objects in foreground (extended vehicles and standing persons) among a sophisticated background (building, tree, bus stop, fence, etc).

To show the performance gain provided by large synthetic aperture, we compare each MIMO-SAR imaging result with its range-angle map counterpart. The range-angle maps - see Fig.~\ref{road_side_exper2}(c), Fig.~\ref{road_side_exper3}(c), and Fig.~\ref{road_side_exper4}(c) - are obtained from the I-Q samples of single frame, using the method in Section \ref{range-angle imaging}.

Based on the analysis of antenna aperture size in Section \ref{sec_simu_parlot}, the synthetic aperture of vehicular-based MIMO-SAR for 3 frames ($\sim$ \SI{100}{ms}) is many times larger than the intrinsic MIMO aperture, resulting in much greater azimuth angle resolution than conventional range-angle imaging.

\subsection{Computation Complexity Analysis}

We compare the computational complexity of different radar imaging algorithms using two metrics: number of float point operations (FLOP) and running time. We run algorithms for the roadside experiment 2 to calculate their FLOPs and averaged running time per frame. 

For MIMO-SAR algorithm, we don't take the complexity of radar odometry into account for a fair comparison. The comparison baselines are the range-angle imaging indicated above, and the backprojection algorithm - a traditional SAR imaging algorithm that only takes the radar I-Q samples of one receiver as input \cite{glob}. We let backprojection algorithm update the whole SAR image every chirp to satisfy the azimuth sampling constraint.

We present the calculated results for three algorithms in Table. \ref{complex_analys}. First, from the comparison between MIMO-SAR and range-angle imaging, we conclude that the performance gain of MIMO-SAR comes at the expense of around 5 times heavier computation load.

Second, it tells that MIMO-SAR algorithm needs almost \textbf{93 times fewer} FLOPs compared to the backprojection algorithm \cite{glob}. While performing on the MATLAB R2019b with a Intel i7-9750H CPU, MIMO-SAR needs \SI{1.17}{s} running time per frame that is \textbf{115 times faster} than backprojection algorithm. Above results confirm our claimed statement - the hierarchical MIMO-SAR algorithm drastically reduces the computational complexity. 

\begin{table}[!t]
\begin{center}
\caption{computational complexity analysis}
\begin{tabular}{lll}
\toprule  
Method & FLOP & Run time per frame\\
\midrule  
MIMO-SAR algorithm & $7.7326 \times 10^7$ & \SI{1.17}{s} \\
Backprojection algorithm \cite{glob} & $7.1532 \times 10^9$ & \SI{134.19}{s} \\
Range-angle imaging & $1.4894 \times 10^7$ & \SI{0.23}{s}\\
\bottomrule 
\label{complex_analys}
\end{tabular}
\end{center}
\vspace{-3mm}
\end{table}

The running time for MIMO-SAR algorithm (\SI{1.17}{s} per frame) can be {\em further reduced} to achieve real-time capability. Since the backprojection processing is suitable for easy parallel implementation \footnote{The calculations are performed pixel-wise on the SAR image, so they can be distributed in GPUs. In addition, the calculations for different snapshots can be done parallelly before the coherent summation.} utilizing GPUs \cite{glob, 8447924}, the calculation time for MIMO-SAR can be efficiently distributed. Besides, some adjustment of parameter values, such as increasing the pixel size $k_{\R{x}}, \, k_{\R{y}}$ of SAR image, and decreasing the side length $\Delta x$, $\Delta y$ of interested region, would reduce the computation load linearly, which allows a trade-off to achieve real-time capability at the cost of image quality degradation.

\subsection{Strengths and Limitations}
To summarize, the proposed MIMO-SAR algorithm enables high azimuth resolution imaging for closely located objects in foreground with much reduced algorithm run time. Experiments in Section~\ref{exper res} show that vehicular-based MIMO-SAR can synthesize a large aperture by exploiting vehicle motion, resulting in better azimuth angle resolution and separation ability (of close targets) than conventional range-angle imaging. Meanwhile, the proposed hierarchical MIMO-SAR algorithm requires \SI{1.17}{s} average run time per frame in Experiment 2, which is 115 times faster compared to its counterpart backprojection algorithm.

\begin{figure}[!t]
    \centering
    \includegraphics[width=3in, trim=2 5 2 2,clip]{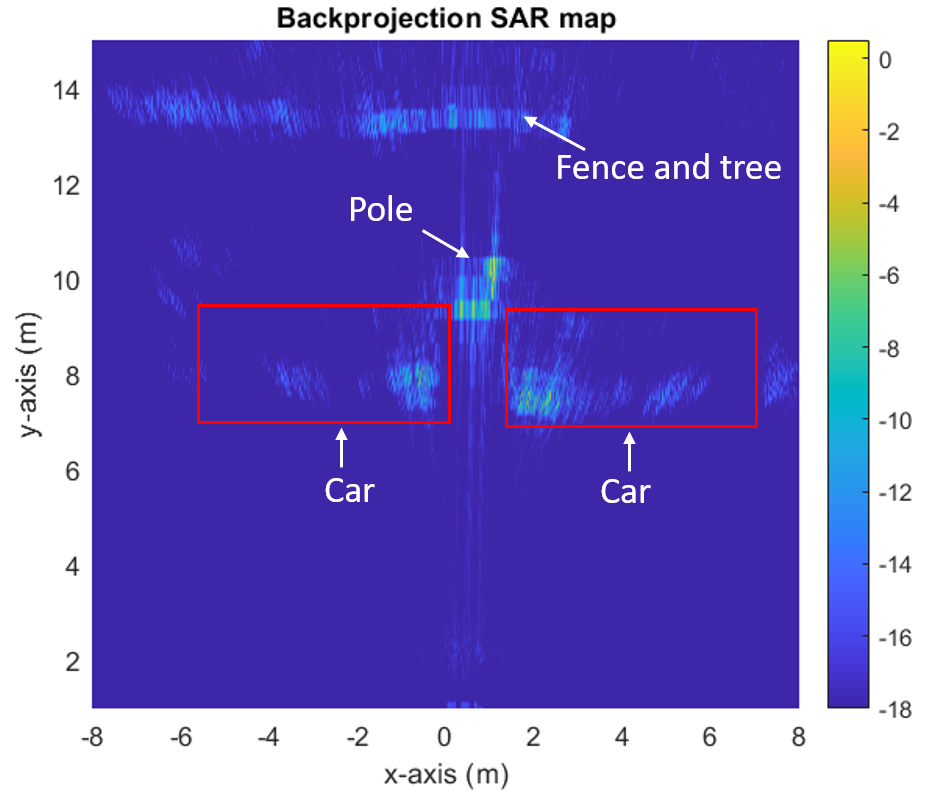}
    \caption{Backprojection SAR imaging result using 3 frames for roadside experiment 2 in Fig.~\ref{road_side_exper3}.}
  \label{bpsar}
\end{figure}

Note that MIMO-SAR preserves the high-resolution imaging quality while reducing computation load, since we exploit all available information on ROIs, i.e., using the virtual aperture created by MIMO, and condensing multiple chirps for each update. To illustrate it, we plot the backprojection SAR imaging result for roadside experiment 2 and show it in Fig.~\ref{bpsar} for comparison with corresponding MIMO-SAR map in Fig.~\ref{road_side_exper3}(b). Results indicate that MIMO-SAR map is enable to visually separate the cars and pole just as the backprojection SAR map, though with 
some losses to the image signal-to-noise ratio (SNR) due to lower PRF.

One limitation of MIMO-SAR algorithm is that it is modeled for stationary targets only and we can not apply it directly for moving targets such as pedestrians and other vehicles. However, it is possible to address this problem by introducing ISAR concepts \cite{9048497, Kulpa2013ExperimentalRO} to MIMO-SAR that incorporates the target motion.

\section{Conclusion}

In this paper, we proposed a new MIMO-SAR algorithm for FMCW automotive radar to get high-resolution imaging. MIMO-SAR incorporates a MIMO processing stage to narrow down the ROI for subsequent finer-resolution SAR processing, which drastically reduces the computation load. Besides, we integrated a radar odometry algorithm to estimate the requisite ego-radar trajectory without relying on additional IMU sensors. We validated the MIMO-SAR by both simulations (via MATLAB) and experiments using a vehicle-mounted side-view radar. For future work, we will continue to explore better way of radar imaging and make the high-resolution radar images apply into practical use.


\bibliographystyle{IEEEtran}
\bibliography{bibtex}

\end{document}